\documentclass[aps,amsmath]{revtex4}
\usepackage{graphicx}

        \newcommand{\be}{\begin{equation}}
        \newcommand{\ee}{\end{equation}}
        \newcommand{\bse}{\begin{subequations}}
        \newcommand{\ese}{\end{subequations}}
        \newcommand{\bea}{\begin{eqnarray}}
        \newcommand{\eea}{\end{eqnarray}}
        \newcommand{\nn}{\nonumber}
        \newcommand{\ban}{\begin{eqnarray*}}
        \newcommand{\ean}{\end{eqnarray*}}
        \newcommand{\half}{\frac{1}{2}}

\renewcommand{\Im}{\mbox{Im}}
\renewcommand{\Re}{\mbox{Re}}

\renewcommand{\d}{{\rm d}}

\begin{document}

\preprint{}

\title{Selfconsistent calculations of 
$\sigma$-meson properties at finite temperature}
\author{Dirk R\"oder}
\email{roeder@th.physik.uni-frankfurt.de}
\affiliation{Institut f\"ur Theoretische Physik,
Johann Wolfgang Goethe-Universit\"at \\
Max-von-Laue-Str.\ 1, D-60438 Frankfurt/Main, Germany}

\begin{abstract}
I study the properties of the scalar $\sigma$-meson [also referred
to as $f_0(600)$] at nonzero temperature
in the $O(N)$-model in the framework of the Cornwall-Jackiw-Tomboulis 
formalism. In the standard Hartree (or large-N) approximation one
only takes into account the double-bubble diagrams in the 
effective potential. I improve
this approximation by additionally taking into account the sunset diagrams,
which leads to 4-momentum dependent real and imaginary parts
of the Dyson-Schwinger equations. By solving these and the equation
for the chiral condensate selfconsistently,
one gets the decay width and the spectral density of the $\sigma$-meson. 
I compare the results in the case with explicit chiral symmetry breaking 
with the chiral limit. I found, that the 4-momentum
dependent real part of the selfenergy do not lead
to qualitative changes in the spectral density.
\end{abstract}

\date{\today}
\maketitle

\section{Introduction}

In the low-energy regime of QCD, chiral symmetry 
is spontaneously broken by a nonvanishing quark condensate 
$\langle\bar{q}q\rangle\sim(300{\rm MeV})^3$. (In nature this symmetry is also
explicitly broken by a nonvanishing quark mass.) At high temperatures,
$T \sim \langle\bar{q}q\rangle$, one expects a ``melting'' of
the condensate and therefore a (partial) restoration of this symmetry. A
major goal in modern hadron physics is to find clear indications for
this chiral restoration. A reasonably way
is to study the properties of strongly interacting matter near
the critical temperature of this transition, e.g. in heavy ion collisions. 

In this work I will focus attention on the sector
of light mesonic degrees of freedom at finite temperature.
An adequate field-theoretical framework for
this is the linear sigma model with $O(N)$ symmetry, which has been
extensively studied in the last decades, cf.
\cite{Gell-Mann:1960np,Baym:1977qb,Bochkarev:1995gi,Roh:1996ek,
Amelino-Camelia:1997dd,Petropoulos:1998gt,Chiku:1998kd,Lenaghan:1999si,
Baacke:2002pi,Patkos:2002xb,Dumitru:2003cf,Roder:2003uz,Andersen:2004ae,
Mocsy:2004ab}
and citations therein.
In this model the fundamental degrees of freedom of QCD
(the quarks and gluons) are integrated out, 
and new (effective) degrees of freedom (the mesons) arise, which 
in the $O(N)$ model are the scalar $\sigma$-meson and the
($N-1$) pseudoscalar pions. These are chiral partners and thus become
degenerate in mass in the chirally restored phase. (For an extension
to finite densities, cf. e.g. 
\cite{Scavenius:2000qd,VicenteVacas:2002se,Cabrera:2005wz}.)

In general, neither QCD nor the $O(N)$ model can be solved
analytically. A standard way to get results
is ordinary perturbation theory which is an expansion
up to a certain order in the coupling constant $g$. Unfortunately, this 
scheme breaks down at nonzero temperature \cite{Dolan:1974qd}.
The reason is that the new energy scale introduced by the temperature
can conspire with the typical momentum scale $p$ of a process
so that $gT/p$ is no longer of order $g$, but can be of order
$1$ \cite{Braaten:1989kk,Braaten:1989mz}. For this reason, one has to find
a proper way to resum whole classes of diagrams of order $gT/p$.

The resummation method I use here is the so called Cornwall-Jackiw-Tomboulis 
(CJT) formalism \cite{Cornwall:1974vz} at finite temperature.
This scheme is equivalent to the
$\Phi$-functional approach of Luttinger and Ward \cite{Luttinger:1960ua} and
Baym \cite{Baym:1962sx}. In the CJT formalism (assuming translation 
invariance),
the Dyson-Schwinger and the condensate equations are given by the minimum
of an effective potential
\begin{equation} \label{eff_cjt}
V[\bar{\sigma},\bar{G}] = U(\bar{\sigma})
+ \frac{1}{2} \,\int_Q \ln \bar{G}^{-1}(Q)
+ \frac{1}{2} \,\int_Q [ G^{-1}(Q,\bar{\sigma})\bar{G}(Q) - 1]
+ V_2[\bar{\sigma}, \bar{G}]
\end{equation}
where $\bar{\sigma}$ and  $\bar{G}$ are the expectation value of the one and
two-point function in the presence of external sources,
$U(\bar{\sigma})$ is the tree-level potential, $G^{-1}$ the inverse tree-level
two-point function, and $V_2[\bar{\sigma},\bar{G}]$
the sum of all two-particle irreducible (2PI) vacuum diagrams with internal 
lines given by $\bar{G}$ (the definition of $\int_Q\dots$
is given at the end of this introduction). Taking into
account {\em all} 2PI diagrams in the
functional $V_2$ would be equivalent to solve the
exact theory, which is impossible. To get numerical
results one has to truncate this sum.

In the standard Hartree (or Hartree-Fock) approximation of the $O(N)$ model,
one only takes into account the double-bubble diagrams, shown
in Figs. \ref{paper1} a, b, c. In these approximations
the selfenergies of the particles are only real valued, therefore no
nonzero decay width effects are included. The difference between
the Hartree and the Hartree-Fock approximations is that 
in the Hartree (or large-$N$) approximation all terms of order $\sim 1/N$
are neglected on the level of the
Dyson-Schwinger and the condensate equations.

In \cite{Roder:2005vt} we presented the so-called improved Hartree-Fock
approximation, which takes additionally into account sunset type diagrams,
shown in Figs. \ref{paper1} d, e. This leads to
4-momentum dependent real and imaginary parts 
in the Dyson-Schwinger equations. Indeed, in that paper, we
neglect the 4-momentum dependent {\em real} parts of the Dyson-Schwinger
equations for simplicity. In the present work, I include them,
although just in the Hartree approximation. This leads to a vanishing imaginary
part of the pion selfenergy (because it is of order $\sim 1/N$),
i.e., the spectral density of the pion is a delta function.
For the $\sigma$-meson the contribution from the $\sigma\to 2\sigma$ decay is
also of order $\sim 1/N$, and vanishes, but the 
contribution from the $\sigma\to 2\pi$ decay remains. Therefore, the spectral 
density has a nonzero width, as expected for the $\sigma$-meson with a very
large vacuum decay width, of $\Gamma_\sigma=(600-1000)~{\rm MeV}$
\cite{PDBook}.
In the following, I call this the improved Hartree approximation. 

The paper is organized as follows. In Sec. II I derive the Dyson-Schwinger
and the condensate equations of the $O(N)$ model in the improved
Hartree approximation. In Sec. III the numerical results
for the $N=4$ case are presented in the chiral limit and in the
case with explicit chiral symmetry breaking for the parameter sets
summarized in Tab.~\ref{parameter}. Section IV concludes
the paper with a summary
and an outlook. The technical derivation of the integrals, which appear
in the Dyson-Schwinger and the condensate equations
are given in the appendix.

I denote 4-vectors by capital letters, $X \equiv(x_0,{\bf x})$, with
${\bf x}$ being a 3-vector of modulus
$x\equiv |{\bf x}|$. 
I use the imaginary time formalism to compute quantities at
finite temperature. Integrals over 4-momentum $K \equiv (k_0, {\bf k})$ are
denoted as
\be
\int_K \, f(K) \equiv T \sum_{n=-\infty}^{\infty}
                       \int \frac{d^{3} k}{(2\pi)^{3}} \,
         f(-i\omega_n,{\bf k}) \;,
\ee
where $T$ is the temperature and
$\omega_n= 2 \pi n T$, $n= 0, \pm 1, \pm 2, \ldots$ 
are the bosonic Matsubara frequencies.
Units are $\hbar=c=k_{B}=1$.  The metric tensor is $g^{\mu \nu}
= {\rm diag}(+,-,-,-)$.

\section{The $O(N)$ model in the improved Hartree approximation}
\label{sectionII}

In this section I discuss the including of the sunset
diagrams in the standard Hartree
approximation. The Lagrangian of the $O(N)$ linear
sigma model is given by 
\be
{\cal L}({\mbox{\boldmath$\phi$}})=
\frac 12\left(\partial_\mu{\mbox{\boldmath$\phi$}}\cdot\partial^\mu
\mbox{\boldmath$\phi$}
-\mu^2\mbox{\boldmath$\phi$}\cdot\mbox{\boldmath$\phi$}\right)
-\frac \lambda N(\mbox{\boldmath$\phi$}\cdot\mbox{\boldmath$\phi$})^2
+H\phi_1\,\,,
\ee
where $\mbox{\boldmath$\phi$}\equiv(\phi_1,\mbox{\boldmath$\pi$})$, 
with the first component $\phi_1$ corresponding to the scalar 
$\sigma$-meson [also referred to as $f_0(600)$]
 and the other components 
$\mbox{\boldmath$\pi$}=(\phi_2,...,\phi_N)$ corresponding to the
pseudoscalar pions. For $H=0$ and $\mu^2>0$, the Lagrangian is invariant
under $O(N)$ rotations of the fields. For negative values of the bare mass
$\mu^2$, the symmetry is spontaneously broken from $O(N)$ to
$O(N-1)$, which leads to $N-1$ Goldstone bosons (the pions). The parameter 
$H$ breaks the $O(N)$ symmetry explicitly and give a
mass to the pions. 

The parameter $H$ is given as a function of the vacuum mass $m_\pi$,
and the vacuum decay constant $f_\pi$ of the pion : $H=m_\pi^2f_\pi$. 
In this work I compare the chiral limit, $m_\pi=0$ and $f_\pi=90~{\rm MeV}$, 
with the case of explicit chiral symmetry breaking, $m_\pi=139.5~{\rm MeV}$ 
and $f_\pi=92.4~{\rm MeV}$. The bare mass $\mu^2$ and the four-point coupling
$\lambda$ depend additionally on the vacuum mass of the $\sigma$-meson
$m_\sigma$: $\mu^2=-(m_\sigma^2-3\,m_\pi^2)/2$, and
$\lambda=N(m_\sigma^2-m_\pi^2)/(8f_\pi^2)$.
The decay width of the $\sigma$-meson in vacuum is very large, 
$\Gamma_\sigma=(600-1000)~{\rm MeV}$, therefore the vacuum mass is not
well defined, $m_\sigma=(400-1200)~{\rm MeV}$ \cite{PDBook}. I compare
the results for $m_\sigma=400,\;600,$ and $800~{\rm MeV}$. The parameter
sets of the model, for $N=4$, are summarized in table \ref{parameter}.
\begin{table}
\begin{center}
\begin{tabular}{|l|l|l|} \hline
$\begin{array}{c}
\mbox{$\sigma$-meson}\\
\mbox{vacuum mass}\\
\end{array}$
& 
$\begin{array}{c}
\mbox{explicit chiral symmetry breaking}\\
m_\pi=139.5\,{\rm MeV},\;\;f_\pi=92.4\,{\rm MeV}\\
H=(121.6\,{\rm MeV})^3
\end{array}$
& $\begin{array}{c}
\mbox{chiral limit}\\
m_\pi=0\,{\rm MeV},\;\;f_\pi=90\,{\rm MeV}\\
H=0
\end{array}$\\ \hline
$ m_{\sigma}=400\,{\rm MeV}$ & $\lambda=8.230 $ & $\lambda=9.877 $\\
                             & $\mu^2= -(225.41\,{\rm MeV})^2$ &
                               $\mu^2= -(282.84\,{\rm MeV})^2$ \\\hline
$ m_{\sigma}=600\,{\rm MeV}$ & $\lambda=19.043 $ & $\lambda=22.222 $\\
                             & $\mu^2= -(388.34\,{\rm MeV})^2$ &
                               $\mu^2= -(424.264\,{\rm MeV})^2$  \\\hline
$ m_{\sigma}=800\,{\rm MeV}$ & $\lambda=36.341 $& $\lambda=39.506 $\\
                             & $\mu^2= -(539.27\,{\rm MeV})^2$ &
                               $\mu^2= -(565.685\,{\rm MeV})^2$ \\\hline
\end{tabular}
\end{center}
\vspace{3mm} 
\caption{ The masses and decay constants at vanishing
temperature and the corresponding parameter sets for the
$O(4)$ linear sigma model for the two symmetry breaking patterns
studied here.}
\label{parameter}
\end{table}

\begin{figure}
\includegraphics[height=3cm]{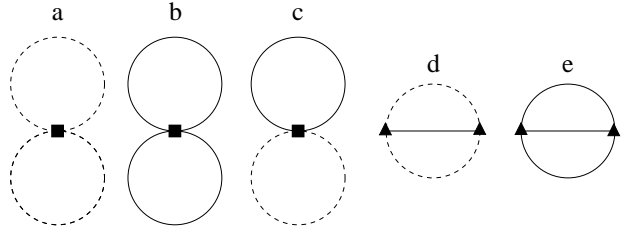}
\caption{The set of two-particle irreducible diagrams included in
the effective potential of the CJT-formalism.
The diagrams a,b, and c are the double-bubble
diagrams, and d and e are the sunset diagrams. Full lines are scalar
particles, dashed lines are pseudoscalar particles. The four-particle vertex 
$\sim \lambda$ is represented by a full squared and the three-particle
vertex $\sim \lambda\sigma$ is represented by a full triangle.}
\label{paper1}
\end{figure}

The effective potential for the $O(N)$ model in the CJT formalism, 
Eq.~(\ref{eff_cjt}), is given by \cite{Roder:2003uz,Lenaghan:1999si}
\bea\label{CJT-potential}
V[\bar{\sigma},\bar{S},\bar{P}]
 & = & \frac 12 \mu^2\bar\sigma^2+\frac{\lambda}{N}\bar\sigma^4
    -   H\bar\sigma\,\,
    +  \half \int_Q \left[\, \ln \bar{S}^{-1}(Q) +
        S^{-1}(Q;\bar{\sigma})\, \bar{S}(Q)-1 \, \right] \nn \\
 & + &  \frac{N-1}{2} \int_Q \, \left[\,
        \ln \bar{P}^{-1}(Q) + P^{-1}(Q;\bar{\sigma})\, \bar{P}(Q)-
        1 \, \right]  +  V_{2}[\bar{\sigma},\bar{S},\bar{P}]\,\, ,
\eea 
where $\bar{\sigma},\bar{S},\bar{P}$ are the expectation values of
the one- and two-point functions in the presence of external
sources (I omit the dependency on the 
pseudoscalar one-point function $\bar{\pi}$,
because the vacuum has even parity, and therefore $\bar\pi=0$). At tree-level
the expectation value of $\bar\sigma$ can be calculated analytically 
\bea\label{vev_tree}
\bar{\sigma}=f_\pi=\sqrt{-\frac{N\mu^2}{4\lambda}}\frac{2}{\sqrt{3}}
             \cos\frac{\theta}{3},\quad
\theta=\arccos\left[
\frac{HN}{8\lambda}\left(-\frac{12\lambda}{N\mu^2}\right)^{3/2}\right].
\eea
The quantities $S^{-1}$ and $P^{-1}$ 
are the inverse tree-level
propagators for scalar and pseudoscalar mesons,
\be
S^{-1}(K;\bar{\sigma}) = -K^{2} + m_\sigma^2(\bar{\sigma})\; , \quad
P^{-1}(K;\bar{\sigma}) = -K^{2} + m_\pi^2 (\bar{\sigma})\; , 
\ee
where the tree-level masses are
\be\label{tree_level_mass}
m_\sigma^2(\bar{\sigma})  =  \mu^{2} + 
        \frac{12\, \lambda}{N}\, \bar{\sigma}^{2}\; ,
 \quad
m_\pi^2(\bar{\sigma})  =     \mu^2 +
        \frac{4\, \lambda}{N}\, \bar{\sigma}^{2}\; . 
\ee
As discussed in the introduction, in the functional $V_2$, additionally
to the three double-bubble diagrams 
shown in Figs.~\ref{paper1} a, b, c,
the two sunset diagrams shown in Figs.~\ref{paper1} d, e,
are taken into account,
\bea \label{V2}
V_{2}[\bar{\sigma},\bar{S},\bar{P}] &\equiv&
        (N+1)(N-1)\, \frac{\lambda}{N} \left[
        \int_Q \, \bar{P}(Q)\right]^{2}
+ 3\, \frac{\lambda}{N}
        \left[ \int_Q \, \bar{S}(Q) \right]^{2}\nn\\
&+& 2\,(N-1) \frac{\lambda}{N}
        \int_Q\, \bar{S}(Q) 
        \int_Q\, \bar{P}(Q)\nn\\
&+& \frac{1}{2} \, 2(N-1)
     \left(\frac{4\lambda \bar\sigma}{N}\right)^2\int_L\int_Q
      \bar{S}(L)\bar{P}(Q)\bar{P}(L+Q)\nn\\
&+&\frac{1}{2}\, 3! \left(\frac{4\lambda \bar\sigma}{N}\right)^2\int_L\int_Q
 \bar{S}(L)\bar{S}(Q)\bar{S}(L+Q).
\eea 
To get the expectation values for the one- and two-point functions in the 
absence of external sources, $\sigma$, ${\cal S}$, and $\cal P$, one has
to find the stationary points of the effective potential 
(\ref{CJT-potential}).
Minimization of the effective potential with respect to the one-point 
function $\delta V/\delta \bar\sigma=0$,
leads to an equation for the scalar condensate $\sigma$, 
\bea
H  &=&  \mu^2 \, \sigma + \frac{4 \lambda}{N}\, \sigma^3
 +  \frac{4\lambda}{N}\, \sigma \int_Q\, 
\left[ 3 \,{{\cal S}}(Q) + (N-1) \, {{\cal P}}(Q)\right] \nn\\
&+&
\left(\frac{4\lambda}{N}\right)^2\sigma
\left[2(N-1)
\int_L\int_Q
{\cal S}(L){\cal P}(Q){\cal P}(L+Q) \right. 
+ \left. 3!\int_L\int_Q
{\cal S}(L){\cal S}(Q){\cal S}(L+Q)\right]\; . \label{dv2ds}
\eea
Using the fact that $\sigma^2\sim N$ [cf. Eq. (\ref{vev_tree})],
this equation becomes in the large-$N$ limit,
\be\label{condensate_largen} 
H  =  \sigma\left\{\mu^2 + \frac{4 \lambda}{N}\left[\sigma^2
 + N\, \int_Q\, {{\cal P}}(Q)\right]\right\}.
\ee
The minimization of the effective potential with respect to
the two-point functions,
$\delta V/\delta \bar{S}=0$ and $\delta V/\delta \bar{P}=0$,
leads to the Dyson-Schwinger equations for the scalar and pseudoscalar
propagators, ${\cal S}$ and ${\cal G}$,
\be\label{schwinger_dyson}
{\cal S}^{-1}(K;\sigma)=S^{-1}(K;\sigma)
+\Sigma(K;\sigma)\;,\quad
{\cal P}^{-1}(K;\sigma)=P^{-1}(K;\sigma)
+\Pi(K;\sigma)\;.
\ee
Here I introduced the selfenergy of the scalar
\bse
\bea\label{self_energy_sigma}
\Sigma(K;\sigma)&\equiv&\frac{4\lambda}{N}\left[ 3\, \int_Q \, {\cal S}(Q)\;
+(N-1)\, \int_Q \, {\cal P}(Q)\right]\; \nn\\
&+&\left(\frac{4\lambda \sigma}{N}\right)^2
\left[2(N-1)\int_Q{\cal P}(K-Q){\cal P}(Q)\; 
+3\cdot 3!\int_Q{\cal S}(K-Q){\cal S}(Q)\right]\;,
\eea
and pseudoscalar fields
\bea\label{self_energy_pion}
\Pi(K;\sigma)&\equiv&\frac{4\lambda}{N}\left[ \int_Q \, {\cal S}(Q)\; 
+(N+1)\,  \int_Q \, {\cal P}(Q)\right]
+\left(\frac{4\lambda \sigma}{N}\right)^2
4\int_Q{\cal P}(K-Q){\cal S}(Q)\,\,.
\eea\ese
In the large-$N$ limit, all $\sigma$-meson tadpoles and a
part of the pion tadpole
($\sim\int{\cal S}$ and $\sim\int{\cal P}$) vanish, 
and only a part of the pion term ($\sim\int{\cal P}{\cal P}$) remains,
\be
\Sigma(K;\sigma)
=\frac{4\lambda}{N}N\, \int_Q \, {\cal P}(Q)\; 
+\left(\frac{4\lambda \sigma}{N}\right)^2
2\,N\int_Q{\cal P}(K-Q){\cal P}(Q)\; ,\quad
\Pi(K;\sigma)=\frac{4\lambda}{N}
N\,  \int_Q \, {\cal P}(Q).
\ee
The tadpole contributions have no imaginary parts, therefore
\be\label{im_largen}
\Im\,\Sigma(K;\sigma)=\left(\frac{4\lambda \sigma}{N}\right)^2
2\,N\,\Im\,\int_Q{\cal P}(K-Q){\cal P}(Q)\;,\quad
\Im\,\Pi=0.
\ee
The whole imaginary part of the $\sigma$-meson selfenergy depends
on the 4-momentum vector, $K=({\bf k},\omega)$, but the real part
can be split into a term which do
not depend on $K$, $[\Re\, \Sigma]_1$, stemming from the tadpole
contribution, and a term which is 4-momentum 
dependent, $[\Re\, \Sigma(K,\sigma)]_2$,
stemming from the sunset diagram,
\bse\be
\Re\, \Sigma(K;\sigma)=[\Re\, \Sigma]_1+[\Re\, \Sigma(K;\sigma)]_2,
\ee
where 
\be\label{re_largen_2}
[\Re\, \Sigma]_1\equiv
\frac{4\lambda}{N}\,N\,\int_Q \, {\cal P}(Q)\;,\quad
[\Re\, \Sigma(K;\sigma)]_2\equiv\left(\frac{4\lambda \sigma}{N}\right)^2
2\,N\,\Re \int_Q{\cal P}(K-Q){\cal P}(Q).
\ee\ese
As shown above, the 4-momentum dependent terms in the pion selfenergy 
vanish in the large-$N$ limit, and the 4-momentum independent term
is the same as for the $\sigma$-meson, $\Re\, \Pi=[\Re\, \Sigma]_1$.

In this work I want to calculate the spectral density of the
$\sigma$-meson $\rho_\sigma$. To this aim, I rewrite
the Dyson-Schwinger equations (\ref{schwinger_dyson}) and the equation 
for the chiral condensate (\ref{condensate_largen}) as functions of the
spectral density of the pion $\rho_\pi$. In the large-$N$ limit, the
imaginary part of the pion vanishes, and therefore the spectral density is
just a delta function 
\be
\label{spectral_density_pion}
\rho_\pi (\omega, k) = \frac{\pi}{\omega_\pi (k)}
 \left\{ \delta [ \omega - \omega_\pi(k)] 
- \delta [ \omega + \omega_\pi(k)] \right\}\;,
\ee
with support on the quasiparticle energy for the pion
$\omega_\pi(k) = \sqrt{k^2 + M_\pi^2(\sigma)}$,
where I defined a 4-momentum independent mass for the pion $M_\pi$,
\be\label{Mpi}
M_\pi^2 (\sigma) \equiv m_\pi^2(\sigma) + \Re\, \Pi\;.
\ee
In the chirally broken phase ($\sigma\ne 0$) the imaginary part of the
$\sigma$-meson is nonzero, therefore the spectral density assumes
the following form: 
\be\label{spectral_density_sigma_1}
\rho_\sigma(\omega, {\bf k})=-\frac{2\,\Im\,\Sigma(\omega, {\bf k};\sigma)}
{[\omega^2-k^2-m_\sigma^2(\sigma)-\Re\,\Sigma(\omega,{\bf k};\sigma)]^2
+[\Im \,\Sigma(\omega, {\bf k};\sigma)]^2}\,\, ,
\ee
where the quasiparticle
energy for the $\sigma$-meson $\omega_\sigma ({\bf k})$, is given
by the solution of
$\omega_\sigma^2({\bf k})-k^2-m_\sigma^2(\sigma)
-\Re\,\Sigma[\omega_\sigma({\bf k}),{\bf k};\sigma]=0$.
In the chirally restored phase ($\sigma=0$) the 4-momentum dependent real and
imaginary parts of the $\sigma$-meson selfenergy vanish, therefore
also the spectral density of the $\sigma$-meson becomes a delta function, 
\be
\label{spectral_density_sigma_2}
\rho_\sigma (\omega, k) = \frac{\pi}{\omega_\sigma (k)}
 \left\{ \delta [ \omega - \omega_\sigma(k)] 
- \delta [ \omega + \omega_\sigma(k)] \right\}\;,
\ee
with support on the quasiparticle energy for the $\sigma$-meson
$\omega_\sigma(k) = \sqrt{k^2 + M_\sigma^2(\sigma)}$,
where I defined a 4-momentum independent mass for 
the $\sigma$-meson $M_\sigma$,
\be\label{Msigma}
M_\sigma^2 (\sigma) \equiv m_\sigma^2(\sigma) + [\Re\, \Sigma]_1\;.
\ee

The selfenergies can be written as functions of the pion spectral density, 
exclusively, which is just a delta function as
discussed above. Note, that for the real
parts, I only consider temperature-dependent contributions and neglect
the (ultraviolet divergent) vacuum parts, which is a simple way to
renormalize the integrals (the imaginary parts are not ultraviolet divergent).
The 4-momentum independent term is just
the standard tadpole integral \cite{Roder:2003uz},
\be
\Re\, \Pi=[\Re\, \Sigma]_1=\frac{1}{2\pi^2}
\int_0^\infty d q\,q^2[\omega_\pi (q)]^{-1}
f[\omega_\pi(q)].
\ee
The derivation of the equations for the 4-momentum dependent terms is
given in the appendix, 
\bse\label{4md}
\bea
\Im\,\Sigma(\omega,k)
&=&\left(\frac{4\lambda \sigma}{N}\right)^2
\frac{N}{8\pi}\frac{1}{k}
\int_0^{\infty}  \, d q \;q\,[\omega_\pi (q)]^{-1}
\Theta\left(|q_0-q|\leq k\leq q_0+q\right)
\nn\\&\times&
\{1+f[\omega_\pi(q_0)]+f[\omega_\pi(q)]\},\\
\mbox{[}\Re\,\Sigma(\omega,k)]_2&=&
\left(\frac{4\lambda \sigma}{N}\right)^2
\frac{N}{8\pi^2}\frac{1}{k}
\mbox{P}\int_0^{\infty}   d q_1 \, q_1 \, d q_2 \, q_2\;
\Theta\left(|q_1-q_2|\leq k\leq q_1+q_2\right)
[\omega_\pi(q_1)\omega_\pi(q_2)]^{-1}\nn\\
&\times& \left\{ 
\frac{f[\omega_\pi(q_1)]+f[\omega_\pi(q_2)]}
{\omega_\pi(q_1)+\omega_\pi(q_2)-\omega}
+\frac{f[\omega_\pi(q_1)]+f[\omega_\pi(q_2)]}
{\omega_\pi(q_1)+\omega_\pi(q_2)+\omega}\right.\nn\\
&&\left.+\frac{-f[\omega_\pi(q_1)]+f[\omega_\pi(q_2)]}
{\omega_\pi(q_1)-\omega_\pi(q_2)-\omega}
+\frac{-f[\omega_\pi(q_1)]+f[\omega_\pi(q_2)]}
{\omega_\pi(q_1)-\omega_\pi(q_2)+\omega}\right\}\;,
\eea\ese
where $f(\omega)\equiv 1/[\exp(\omega/T)-1]$ is the Bose-Einstein
distribution function, $q_0\equiv\sqrt{[\omega-\omega_\pi(q)]^2-M_\pi^2}$,
and $\mbox{P}\int\dots$ denotes the principal value of the integral. Note that 
$\Theta\left(|q_1-q_2|\leq k\leq q_1+q_2\right)/k=2\delta(q_1-q_2)$ 
in the limit $k\to 0$, which can be used to perform the $q$-integration,
cf. Eqs.~(\ref{im2k0}), and (\ref{re2k0}). An appropriate way to perform
the principal value numerically is discussed in the appendix,
cf. Eqs.~(\ref{re2k}), and (\ref{re2k0}).

The spectral densities have to obey a sum rule \cite{leBellac},
$\int_{-\infty}^\infty \d \omega \, \omega\,
\rho_{\sigma,\pi}(\omega,{\bf k}) = 2\pi\;$,
which is {\it{a priori}} fulfilled for the pion spectral density
(\ref{spectral_density_pion}), because of the normalization.
As mentioned above, this is not the case for the 
$\sigma$-meson in the chirally broken phase. The main reason for a
possible violation of the sum rule is due to
neglecting terms of the order $\sim 1/N$ in the selfenergy,
which leads to a loss of spectral strength. 
(I found that the inclusion of the 
4-momentum dependent real part $[\Re\,\Sigma]_2$, is close to negligible
for the validity of the sum rule.)
Other possible problems arise from the numerical realisation of
the spectral density on a finite energy-momentum grid, which 
I minimized by using a large and fine grid.
As discussed in \cite{Roder:2005vt}, if the sum rule is not fulfilled, 
I use the following way to restore it. If the imaginary part
of the $\sigma$-meson selfenergy is very small (i.e. the spectral density
is narrow), I add a numerical realisation of a delta function
$\delta_{\rm num}$ to the spectral density:
$\rho_\sigma'(\omega,k)\rightarrow\rho_\sigma(\omega,k)
+c_1\cdot\delta_{\rm num}[\omega-\omega_\sigma(k)]$. On the other hand,
if the imaginary part of the selfenergy is large enough (compared
with the lattice spacing) I just multiply the
spectral density by a factor:
$\rho_\sigma'(\omega,k)\rightarrow c_2\cdot\rho_\sigma(\omega,k)$.
The constants $c_1$ and $c_2$ are adjusted in that way, that 
$\rho_\sigma'$ fullfills the sum-rule on the (finite) energy-momentum grid.
(I checked, by comparing the results with the case $c_2=1$, that this
procedure does not lead to major quantitative changes.)

The numerically scheme for the improved Hartree approximation is the
following. At first, one solves the standard Hartree approximation,
i.e., the condensate
and the Dyson-Schwinger equations, Eqs.~(\ref{condensate_largen})
and (\ref{schwinger_dyson}), with $[\Re\,\Sigma]_2=[\Im\,\Sigma]_2=0$, to
get the chiral condensate and
the 4-momentum independent mass of the pion, $\sigma$ and $M_\pi$.
With these results, one calculates the
4-momentum dependent real and imaginary parts (\ref{4md}) 
of the $\sigma$-meson selfenergy, and finally 
the spectral density, Eq.~(\ref{spectral_density_sigma_1}) or
(\ref{spectral_density_sigma_2}). Then, the decay width 
of the $\sigma$-meson $\Gamma_\sigma$ is defined as 
\cite{Weldon:1983jn,leBellac}:
$\Gamma_\sigma (k) = \Im\,\Sigma[\omega_\sigma(k),k;\sigma]/\omega_\sigma(k)$,
where $\omega_\sigma(k)$ is given 
by the solution of $\omega_\sigma^2(k)-k^2-m_\sigma^2(\sigma)
-\Re\,\Sigma[\omega_\sigma(k),k;\sigma]=0$.

\section{Results}\label{results}
In this section I present the numerical results for the linear
sigma model with $O(4)$ symmetry in the improved Hartree approximation,
for the parameter sets given in table \ref{parameter}. 

\subsection{The mass}\label{res_1}

\begin{figure}
\includegraphics[height=12cm]
{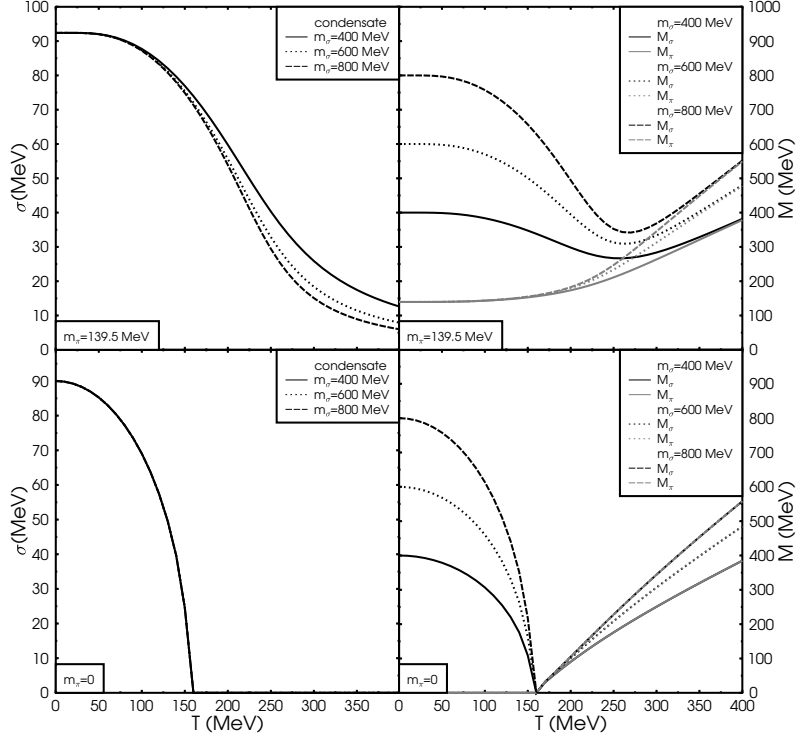}
\caption{The chiral condensate $\sigma$ (left column) 
and the 4-momentum independent mass of the $\sigma$-meson and the pion,
$M_\sigma$ and $M_\pi$ (right column),
as function of temperature $T$ and $\sigma$-meson vacuum mass $m_\sigma$. 
In the upper row are shown the results with explicit chiral symmetry breaking 
and in the lower row the results in the chiral limit.}
\label{Hartree_values}
\end{figure}
\begin{figure}
\includegraphics[height=8cm]
{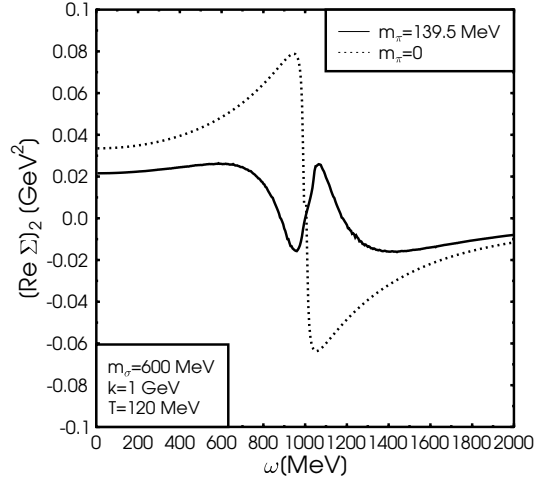}
\caption{The 4-momentum dependent real part
of the $\sigma$-meson selfenergy $[\Re\,\Sigma]_2$, at a
fixed momentum of $k=1$ GeV, for the case with  explicit chiral
symmetry breaking (full line) and the corresponding result
for the chiral limit (dotted line) at a temperature of $T=120$ MeV.}
\label{re_t100_k1000}
\end{figure}

I start the discussion of the results 
with the chiral condensate $\sigma$ and the
4-momentum independent 
mass of the $\sigma$-meson and the pion, $M_\sigma$ and $M_\pi$,
as defined in Eqs.~(\ref{Msigma}) and (\ref{Mpi}).
Note, that the 4-momentum independent mass $M_\sigma$ is
generally {\it not} the physical (pole) mass of the $\sigma$-meson, because the 
pole of the $\sigma$-meson spectral density is additionally shifted
by the 4-momentum dependent part of the selfenergy $[\Re\,\Sigma(K,\sigma)]_2$,
cf. Fig.~\ref{re_t100_k1000}.

In the upper row of Fig.~\ref{Hartree_values} the results for the chiral
condensate $\sigma$ (left column) and the 4-momentum independent
mass for the pion 
and the $\sigma$-meson, $M_\pi$ and $M_\sigma$
(right column), are shown as functions of $T$
and $m_\sigma$, in the case of explicit chiral symmetry 
breaking. The results show
the behaviour of a crossover transition. Neither
the chiral condensate nor the 4-momentum independent 
mass of the scalar particle $M_\sigma$
become equal zero, even for high temperatures. Thus, $M_\pi$
and $M_\sigma$, become only approximatively degenerate for high temperatures.
In the lower row of Fig.~\ref{Hartree_values} the corresponding results for 
the chiral limit are presented. The results show the behaviour of 
a second-order phase transition, which agrees with the predictions 
made by Pisarski and Wilczek \cite{Pisarski:1984ms}.
The condensate $\sigma$ becomes equal zero at a
critical temperature $T_\chi$, therefore $M_\sigma=M_\pi$ 
for $T\ge T_\chi$. The condensate (accordingly $T_\chi$) does not
depend on the vacuum mass $m_\sigma$, which can be understood 
as a consequence of the condensate equation (\ref{condensate_largen}) 
in the chiral limit ($H=0$),
\be
0  =  \mu^2 + \frac{4 \lambda}{N}\left[\sigma^2
 + N\, \int_Q\, {{\cal P}}(Q)\right].
\ee
For $m_\pi=0$ the integral can be performed 
analytically \cite{Lenaghan:1999si},
\be
0  =- \frac{m_{\sigma}^2}{2} + \frac{4 \lambda}{N}\sigma^2
 + 4\lambda \frac{T^2}{12}
\Longrightarrow\sigma=\sqrt{f_\pi^2-\frac{T^2N}{12}}
\ee
[cf. Eq.~(\ref{tree_level_mass})],
and the critical temperature 
is given by $T_\chi=\sqrt{12/N}f_\pi\approx 160~{\rm MeV}$ 
for the $O(4)$ model, which agrees with the numerical results.

The 4-momentum dependent real part of the $\sigma$-meson selfenergy
$[\Re\,\Sigma]_2$ is shown in Fig.~\ref{re_t100_k1000} at fixed momentum
of $k=1$ GeV as a function of energy. It 
is larger in the chiral limit as in the case with explicit chiral
symmetry breaking, because 
$[\Re\,\Sigma]_2\sim \lambda^2\sim (m_\sigma^2-m_\pi^2)^2$ 
is maximal for $m_\pi^2=0$. Note that $[\Re\Sigma]_2$ is
small compared to the (squared) 4-momentum independent mass
of the $\sigma$-meson, shown in Fig.~\ref{Hartree_values}.

\subsection{The decay width}\label{res_2}
After calculating the (4-momentum dependent)
imaginary part of the $\sigma$-meson selfenergy
(\ref{im_largen}), the decay width $\Gamma_\sigma$ 
is calculated at the quasiparticle energy:
$\Gamma_\sigma (k) = 
\Im\,\Sigma[\omega_\sigma(k),k;\sigma]/\omega_\sigma(k)$,
where $\omega_\sigma(k)$ is given 
by the solution of $\omega_\sigma^2(k)-k^2-m_\sigma^2(\sigma)
-\Re\,\Sigma[\omega_\sigma(k),k;\sigma]=0$.
The imaginary part of the selfenergy, and
therefore the decay width, of the pion is of order $\sim 1/N$,
and therefore neglected.

\begin{figure}
\includegraphics[height=8cm]
{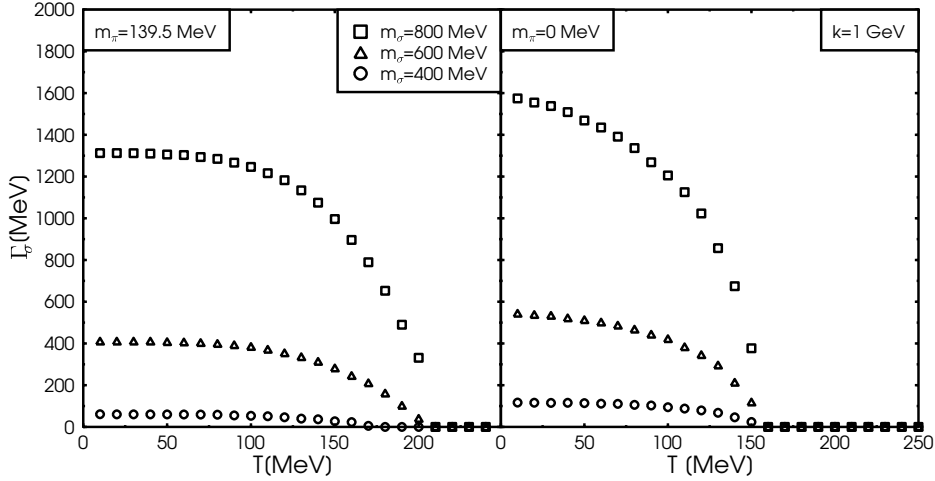}
\caption{The decay width of the $\sigma$-meson $\Gamma_\sigma$,
as a function of temperature $T$ and $\sigma$-meson vacuum mass $m_\sigma$,
at momentum $k=1$ GeV, in the case with explicit chiral symmetry breaking 
(left) and in the chiral limit (right).}
\label{decay_width}

\end{figure}
In Fig.~\ref{decay_width} the decay width of the $\sigma$-meson
$\Gamma_\sigma$ is shown as a function of temperature $T$.
The qualitative behaviour is similar in all cases, but the decay width 
is larger in the chiral
limit compared to the case with explicit chiral symmetry breaking, because
$\Im\,\Sigma\sim\lambda^2\sim (m_\sigma^2-m_\pi^2)^2$ is maximal
for $m_\pi^2=0$, cf. above discussion of $[\Re\,\Sigma]_2$.
The decay width is a strictly monotonic decreasing function with temperature,
and becomes approximatively zero (equal zero)
for temperatures larger than $\sim 200$ MeV 
($T_\chi\approx 160$ MeV) in the case of explicit chiral symmetry breaking 
(the chiral limit). This is a consequence of the (partial) restoration 
of the chiral symmetry, the masses of the chiral partners
become (approximatively) degenerate in mass, and therefore the
phase space of the $\sigma\to 2 \pi$ decay is squeezed. 
The dependence on the vacuum mass of the $\sigma$-meson is significant. 
The reason for this is that $\Gamma_\sigma\sim\Im\,\Sigma\sim m_\sigma^4$,
which agrees reasonably with the results, 
$\Im\,\Sigma=1:3:16$ for $m_\sigma=400:600:800$ MeV.

\begin{figure}
\includegraphics[height=8cm]
{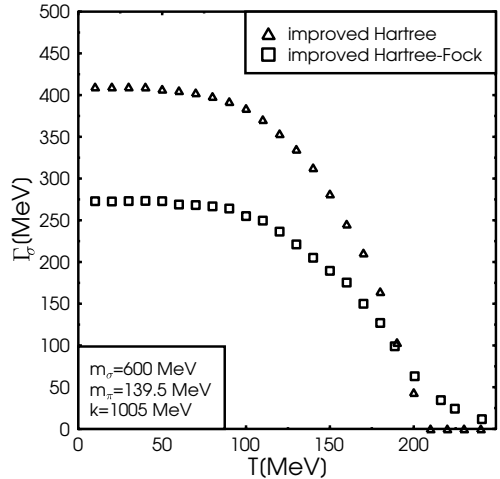}
\caption{The decay width of the $\sigma$-meson $\Gamma_\sigma$,
in the improved Hartree
(triangles) and in the improved Hartree-Fock (squares) approximation, 
as a function of temperature $T$, at momentum $k=1005$ MeV and 
$\sigma$-meson vacuum mass of $m_\sigma=600$ MeV,
in the case with explicit chiral symmetry breaking.}
\label{decay_width_cmp_hf_largen}
\end{figure}
In Fig.~\ref{decay_width_cmp_hf_largen}
the decay width of the $\sigma$-meson in the improved Hartree
(triangles) is compared with the improved Hartree-Fock \cite{Roder:2005vt} 
(squares) approximation. 
The main difference stems from the combinatorial factor 
in front of the two-pion term ($\sim\int{\cal P}{\cal P}$) in 
the imaginary part of the selfenergy [cf. Eqs.~(\ref{self_energy_sigma})
with (\ref{im_largen})]. A part of this contribution is
of order $\sim 1/N$, and neglected in the Hartree approximation. Therefore,
this factor is $2\cdot N=8$ in the improved Hartree, and $2\cdot(N-1)=6$
in the improved Hartree-Fock approximation, which would lead to 
a decay width of an factor $\approx 1.33$ larger
in the improved Hartree approximation.
The remaining difference, shown in the plot,
comes from the $\sigma$-meson term
($\sim3\cdot 3!\int{\cal S}{\cal S}$) in Eq. (\ref{im_largen}),
which vanishes in the large-$N$ limit.

\subsection{The spectral density}\label{res_3}
Finally, after solving the coupled condensate
and Dyson-Schwinger equations selfconsistently,
Eqs.~(\ref{condensate_largen})
and (\ref{schwinger_dyson}), the spectral density of the $\sigma$-meson
$\rho_\sigma$
is given in the chiral broken phase ($\sigma\neq 0$) by 
Eq.~(\ref{spectral_density_sigma_1}),
and in the restored phase ($\sigma=0$) by
Eq.~(\ref{spectral_density_sigma_2}).

\begin{figure}
\includegraphics[height=8cm]
{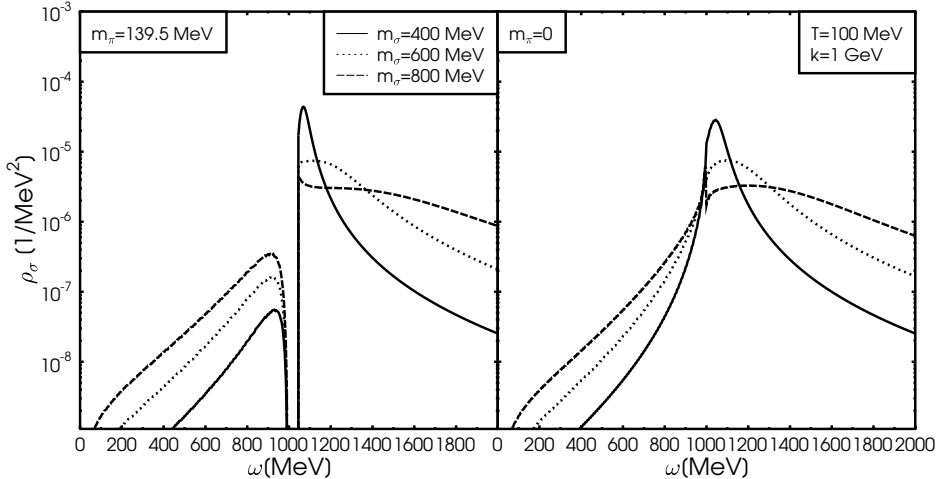}
\caption{The spectral density of the $\sigma$-meson $\rho_\sigma$, 
as a function of energy $\omega$, and $\sigma$-meson vacuum mass $m_\sigma$,
at a momentum of $k=1$ GeV and a temperature of $T=100$ MeV,
in the case with explicit chiral symmetry
breaking (left) and the chiral limit (right).}
\label{rho_sigma_cutp_xs_cl}
\end{figure}
In Fig.~\ref{rho_sigma_cutp_xs_cl} the spectral density
is compared for all possible parameter sets (cf. table I)
at a fixed momentum of $k=1$ GeV.
The spectral density does not exhibit a 
pronounced peak at the quasiparticle energy
$\omega_\sigma(k)$, because this
energy is large enough for the
decay into two pions. As discussed above, the decay width
of this process is large and becomes even larger
for increasing $m_\sigma$, as shown in the plot.
A remarkable difference between the chiral limit (right) and the
case with explicit chiral symmetry breaking (left) is, that
$\rho_\sigma\approx 0$ around $\omega\approx 1$ GeV in
the case with explicit symmetry breaking but not in the chiral limit. The
reason is, that the threshold energy for $\sigma\rightarrow 2\pi$ is 
$\omega=2M_\pi$, and therefore zero in the chiral limit.
For $\omega<k$ the $\sigma$-meson is Landau-damped.

\begin{figure}
\includegraphics[height=8cm]
{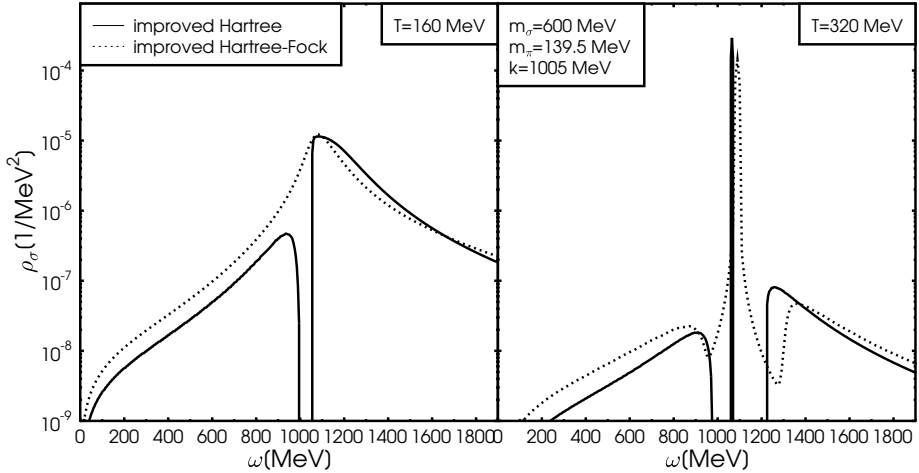}
\caption{The spectral density of the $\sigma$-meson $\rho_\sigma$,
as a function of energy $\omega$, at a momentum of $k=1$ GeV, 
and a vacuum $\sigma$-meson mass of $m_\sigma=600 $MeV, in the
case with explicit chiral symmetry breaking. The results are shown for 
the improved Hartree (full line) and the improved Hartree-Fock (dotted line)
approximation at temperatures of $T=160$ MeV (left), 
and 320 MeV (right).}
\label{rho_sigma_vgl_hfln}
\end{figure}

To illustrate the effects of neglecting terms of order
$\sim 1/N$ in the improved Hartree approximation, I compare in
Fig.~\ref{rho_sigma_vgl_hfln} the results with the improved
Hartree-Fock \cite{Roder:2005vt} approximation,
in the low-temperature regime at $T=160$ MeV (left), and in the
high-temperature regime $T=320$ MeV (right).
As show, the terms of order $\sim 1/N$
(included in the improved Hartree-Fock
approximation) leads to a nonzero decay
width of the pion, and therefore to a 
washed-out $\sigma$-meson spectral density.
One aim of this work was to study the influence of the
4-momentum dependent {\it real} parts of
the selfenergy. I found, that they does not
lead to qualitative changes of the spectral density.
To get an quantitative estimate, I consider the relative change between 
the spectral density with and without this real part, averaged over
the energy-momentum grid, which is rather small,
e.g. for the results shown in Fig. \ref{rho_sigma_vgl_hfln}:
$5.61\pm 2.90\%$ for $T=160~{\rm MeV}$,
and $0.44\pm 0.01\%$ for $T=320~{\rm MeV}$.

\section{Conclusions}\label{conclusion}
In this paper I systematically improved the standard Hartree (or large-$N$)
approximation of the $O(N)$ linear sigma model by taking into
account, additionally to the double-bubble diagrams 
(shown in Figs. \ref{paper1} a, b, c),
the sunset diagrams (shown in Figs. \ref{paper1} d, e)
in the 2PI effective potential of the CJT formalism. 
This leads to 4-momentum dependent real and imaginary parts
of the Dyson-Schwinger equations. 
I solve these and the equation of the condensate
selfconsistently to get the decay width 
and the spectral density of the $\sigma$-meson, for the 
parameter sets summarized in table I.

First, I presented the results for the real parts
of the Dyson-Schwinger, and the condensate equations.
The 4-momentum independent parts, i.e., the 4-momentum independent
masses and the chiral condensate exhibit a crossover transition
in the case with explicit chiral symmetry breaking and 
a second order phase transition in the chiral limit,
as predicted by Pisarski and Wilczek \cite{Pisarski:1984ms}. 
The 4-momentum dependent real parts are small compared to
the (squared) 4-momentum independent masses.

The decay width, presented in the second part of the 
results, show qualitatively the same behaviour in all cases.
It is a strictly monotonic decreasing function with temperature,
and becomes approximatively zero (equal zero)
for temperatures larger than $\sim 200$ MeV 
($T_\chi\approx 160$ MeV) in the case of explicit chiral symmetry breaking 
(the chiral limit). This is a consequence of the (partial) restoration 
of the chiral symmetry, the masses of the chiral partners
becomes (approximatively) degenerate in mass, and therefore the
phase space of the $\sigma\to 2 \pi$ decay is squeezed. 
The dependence on the vacuum mass of the $\sigma$-meson is significant,
because of $\Gamma_\sigma\sim\Im\,\Sigma\sim m_\sigma^4$.

Finally, the behaviour of the decay width and the real parts are 
reflected in the spectral density. In the low-temperature regime,
the spectral density is a very broad 
function in energy (and becomes even broader
for larger vacuum mass $m_\sigma$)
due to the $\sigma\to 2\pi$ decay and becomes more and more
a delta function for increasing temperatures. The 4-momentum
dependent real part of the selfenergy do not lead
to qualitative changes.

The next step is to include also terms of order
$\sim 1/N$, which would lead to a nonvanishing decay width
of the pion and therefore to washed-out spectral
densities. Other possible future 
projects are the inclusion of more degrees
of freedom, for instance baryons \cite{Beckmann:2005nm}, and vector mesons
\cite{vanHees:2000bp,Ruppert:2004yg,strueberdipl}. 
The latter are of particular importance, since
in-medium changes in the spectral properties of vector mesons 
are reflected in the dilepton spectrum \cite{Ruppert:2004yg}
which, in turn, is experimentally observable in heavy-ion collisions
at GSI-SIS, CERN-SPS and BNL-RHIC energies.

\section*{Acknowledgements}
I would like to thank J. Schaffner-Bielich,
L. Tolos, C. Beckmann, J. Ruppert,
and especially D. H. Rischke for fruitful discussions.
The Center of Scientific Computing of the University of Frankfurt
provided computer time.

\section*{Appendix: The 4-momentum dependent parts of the selfenergy}
In this section I derive the equations for the 4-momentum dependent
imaginary and real part of the selfenergy (stemming from the sunset 
diagram). As discussed in \cite{Roder:2005vt}
this diagram can be expressed as a function of the spectral density
$\rho_\pi$ in the imaginary time formalism
\bea\label{cut_sunset_1}
\Sigma(-i\omega_m,k)&=&
\left(\frac{4\lambda \sigma}{N}\right)^2
\frac{2N}{(2\pi)^4}\frac{1}{k}
\int_{-\infty}^{\infty} d \omega_1\,d \omega_2
\frac{1+f(\omega_1)+f(\omega_2)}{i\omega_m+\omega_1+\omega_2}\nn \\
&\times& 
\int_0^{\infty}   d q_1 \, q_1 \,d q_2 \, q_2\;
\Theta\left(|q_1-q_2|\leq k\leq q_1+q_2\right)
 \rho_\pi(\omega_1,q_1)\rho_\pi(\omega_2,q_2),
\eea
where $f(\omega)=1/[\exp(\omega/T)-1]$ is the Bose-Einstein 
distribution function, and $\omega_m=2\pi m T$ are the Matsubara
frequencies. To extract the imaginary part, one uses the Dirac identity, 
$\Im 1/(x+i\epsilon)=-\pi\delta(x)$, and analytic continuation,
$-i\omega_m\rightarrow \omega+i\epsilon$, for the retarded self-energy,
\bea
\Im\,\Sigma(\omega,k)&=&
\left(\frac{4\lambda \sigma}{N}\right)^2
\frac{N}{2(2\pi)^3}\frac{1}{k}
\int_{-\infty}^{\infty} d \omega_1\,d \omega_2
[1+f(\omega_1)+f(\omega_2)]\,\delta(\omega-\omega_1-\omega_2)\nn \\
&\times& 
\int_0^{\infty}   d q_1 \, q_1 \,d q_2 \, q_2\;
\Theta\left(|q_1-q_2|\leq k\leq q_1+q_2\right)
\rho_\pi(\omega_1,q_1)\rho_\pi(\omega_2,q_2).
\eea
As discussed in Sec.~\ref{sectionII} in the improved Hartree approximation 
of the $O(N)$ model the spectral density of the pion 
is just a delta function (\ref{spectral_density_pion}), 
\bea
\Im\,\Sigma(\omega,k)&=&
\left(\frac{4\lambda \sigma}{N}\right)^2
\frac{N}{8\pi}\frac{1}{k}
\int_{-\infty}^{\infty} d \omega_1\, d \omega_2
[1+f(\omega_1)+f(\omega_2)]\,\delta(\omega-\omega_1-\omega_2)\nn \\
&\times& 
\int_0^{\infty}   d q_1 \, q_1 \,d q_2 \, q_2\;
\Theta\left(|q_1-q_2|\leq k\leq q_1+q_2\right)
[\omega_\pi (q_1)\omega_\pi (q_2)]^{-1}\nn\\
&\times&
 \left\{ 
\delta [ \omega_1 - \omega_\pi(q_1)] 
\,\delta [ \omega_2 - \omega_\pi(q_2)]+
\delta [ \omega_1 + \omega_\pi(q_1)]
\,\delta [ \omega_2 + \omega_\pi(q_2)]\right.\nn\\
&&-\left.
\delta [ \omega_1 - \omega_\pi(q_1)] 
\,\delta [ \omega_2 + \omega_\pi(q_2)]-
\delta [ \omega_1 + \omega_\pi(q_1)] 
\,\delta [ \omega_2 - \omega_\pi(q_2)]
 \right\}\;,
\eea
where $\omega_\pi(q)=\sqrt{q^2+M_\pi^2}$ is the quasiparticle energy of the
pion. To simplify this expression one uses the 
$\delta(\omega-\omega_1-\omega_2)$ function to
 carry out the $\omega_2$-integration, 
\bea
\Im\,\Sigma(\omega,k)
&=&
\left(\frac{4\lambda \sigma}{N}\right)^2
\frac{N}{8\pi}\frac{1}{k}
\int_{-\infty}^{\infty} d \omega_1
[1+f(\omega_1)+f(\omega-\omega_1)]\nn\\
&\times&\int_0^{\infty}   d q_1 \, q_1 \, d q_2 \, q_2\;
\Theta\left(|q_1-q_2|\leq k\leq q_1+q_2\right)
[\omega_\pi (q_1)\omega_\pi (q_2)]^{-1}\nn\\
&\times& 
 \left\{ 
\delta [ \omega_1 - \omega_\pi(q_1)] 
\,\delta [ \omega-\omega_1 - \omega_\pi(q_2)]+
\delta [ \omega_1 + \omega_\pi(q_1)]
\,\delta [ \omega-\omega_1 + \omega_\pi(q_2)]\right.\nn\\
&&-\left.
\delta [ \omega_1 - \omega_\pi(q_1)] 
\,\delta [ \omega-\omega_1 + \omega_\pi(q_2)]-
\delta [ \omega_1 + \omega_\pi(q_1)] 
\,\delta [ \omega-\omega_1 - \omega_\pi(q_2)] \right\}\;,
\eea
and the $\omega_1$-integration with the help of the
$\delta[\omega_1\pm \omega_\pi(q_1)]$-functions
\bea\label{cut_sunset_im_1}
\Im\,\Sigma(\omega,k)
&=&\left(\frac{4\lambda \sigma}{N}\right)^2
\frac{N}{8\pi}\frac{1}{k}
\int_0^{\infty}   d q_1 \, q_1 \, d q_2 \, q_2\;
\Theta\left(|q_1-q_2|\leq k\leq q_1+q_2\right)
[\omega_\pi (q_1)\omega_\pi (q_2)]^{-1}\nn\\
&\times&(\,
\{1+f[\omega_\pi(q_1)]+f[\omega-\omega_\pi(q_1)]\}
\,\delta [ \omega-\omega_\pi(q_1) - \omega_\pi(q_2)]\nn\\
&&+\{1+f[-\omega_\pi(q_1)]+f[\omega+\omega_\pi(q_1)]\}
\,\delta [ \omega+\omega_\pi(q_1) + \omega_\pi(q_2)]\nn\\
&&-\{1+f[\omega_\pi(q_1)]+f[\omega-\omega_\pi(q_1)]\}
\,\delta [ \omega-\omega_\pi(q_1) + \omega_\pi(q_2)]\nn\\
&&-\{1+f[-\omega_\pi(q_1)]+f[\omega+\omega_\pi(q_1)]\}
\,\delta [ \omega+\omega_\pi(q_1) - \omega_\pi(q_2)])\;.
\eea
The $\delta[\omega+\omega_\pi(q_1)+\omega_\pi(q_2)]$ function has no support,
because $\omega>0$ and $\omega_\pi(q)>0$. Note that evaluating the 
$\delta [ \omega-\omega_\pi(q_1) + \omega_\pi(q_2)]$
and the $\delta [ \omega+\omega_\pi(q_1) - \omega_\pi(q_2)]$ function,
the latter two terms cancels each other. The remaining
delta function can be transformed to a delta function in momentum-space,
$\delta [ \omega-\omega_\pi(q_1) - \omega_\pi(q_2)]=
|\omega_\pi(q_0)/q_0|\,[\delta(q_1-q_0)+\delta(q_1+q_0)]$
where $q_0\equiv\sqrt{[\omega-\omega_\pi(q_2)]^2-M_\pi^2}$,
is the root of the argument of the delta function. Note that 
in Eq.~(\ref{cut_sunset_im_1}) $q_1>0$ and therefore there is no support 
of the $\delta(q_1+q_0)$ function. Carrying out the the $q_1$-integration, 
using $\omega-\omega_\pi(q_0)=\omega_\pi(q_1)$, and relabelling
$q\equiv q_2$, one gets
\bea
\Im\,\Sigma(\omega,k)
&=&\left(\frac{4\lambda \sigma}{N}\right)^2
\frac{N}{8\pi}\frac{1}{k}
\int_0^{\infty}  \, d q \;q\,[\omega_\pi (q)]^{-1}
\Theta\left(|q_0-q|\leq k\leq q_0+q\right)
\nn\\&\times&
\{1+f[\omega_\pi(q_0)]+f[\omega_\pi(q)]\}.
\eea
To calculate the limit $k\to 0$, best one starts with
Eq.~(\ref{cut_sunset_im_1}), uses the transformation
$1+f[\omega_\pi(q_1)]+f[\omega-\omega_\pi(q_1)]
=[1-\exp(-\omega/T)]\{1+f[\omega_\pi(q_1)]\}
\{1+f[\omega-\omega_\pi(q_1)]\}$,
the identity 
\bea\label{theta_delta}
\lim_{k\to 0}\frac{\Theta\left(|q_1-q_2|\leq k\leq q_1+q_2\right)}{k}
=2\,\delta(q_1-q_2)
\eea
(to perform the $q_1$-integration), and relabels $q\equiv q_2$, to get
\bea
\Im\,\Sigma(\omega,0)
&=&
\left(\frac{4\lambda \sigma}{N}\right)^2
\frac{N}{4\pi}\left[1-\exp\left(-\frac{\omega}{T}\right)\right]
\int_0^{\infty}   d q \, q^2\,\delta [\omega-2\omega_\pi(q)]
[\omega^2_\pi (q)]^{-1}\nn\\
&\times&\{1+f[\omega_\pi(q)]\}\{1+f[\omega-\omega_\pi(q)]\}.
\eea
As discussed in detail in appendix E of Ref. \cite{Rischke:1998qy},
this expression can be calculated analytically, using the Spence integral,
\be\label{im2k0}
\Im\,\Sigma(\omega,0)=
\left(\frac{4\lambda \sigma}{N}\right)^2 N
\frac{1}{8\pi}\sqrt{1-\frac{4M_\pi^2}{\omega^2}}\coth\frac{\omega}{4T}\;.
\ee
Note that in \cite{Rischke:1998qy} the calculation is performed for the
special case $\omega=m_\sigma$, but this can be generalised without
further problems.

To calculate the real part of Eq.(\ref{cut_sunset_1}),
one has to integrate over the principal value of 
$1/(\omega_1+\omega_2-\omega)$, denoted by $\mbox{P}\int\dots$,
\bea
[\Re\,\Sigma(\omega,k)]_2&=&
\left(\frac{4\lambda \sigma}{N}\right)^2
\frac{2N}{(2\pi)^4}\frac{1}{k}
\mbox{P}\int_{-\infty}^{\infty} d \omega_1\,d \omega_2
\frac{1+f(\omega_1)+f(\omega_2)}
{\omega_1+\omega_2-\omega}\nn \\
&\times& 
\int_0^{\infty}   d q_1 \, q_1 \,d q_2 \, q_2\;
\Theta\left(|q_1-q_2|\leq k\leq q_1+q_2\right)
\rho_\pi(\omega_1,q_1)\rho_\pi(\omega_2,q_2).
\eea
Again, one first uses the fact that the spectral density 
of the pion is just a delta function, which can be used to perform the
$\omega_1$- and the $\omega_2$-integration. Using trivial renormalisation,
i.e., neglecting the divergent parts leads to
\bea
[\Re\,\Sigma(\omega,k)]_2&=&
\left(\frac{4\lambda \sigma}{N}\right)^2
\frac{N}{8\pi^2}\frac{1}{k}
\mbox{P}\int_0^{\infty}   d q_1 \, q_1 \, d q_2 \, q_2\;
\Theta\left(|q_1-q_2|\leq k\leq q_1+q_2\right)
[\omega_\pi(q_1)\omega_\pi(q_2)]^{-1}\nn\\
&\times& \left\{ 
\frac{f[\omega_\pi(q_1)]+f[\omega_\pi(q_2)]}
{\omega_\pi(q_1)+\omega_\pi(q_2)-\omega}
+\frac{f[\omega_\pi(q_1)]+f[\omega_\pi(q_2)]}
{\omega_\pi(q_1)+\omega_\pi(q_2)+\omega}\right.\nn\\
&&\left.+\frac{-f[\omega_\pi(q_1)]+f[\omega_\pi(q_2)]}
{\omega_\pi(q_1)-\omega_\pi(q_2)-\omega}
+\frac{-f[\omega_\pi(q_1)]+f[\omega_\pi(q_2)]}
{\omega_\pi(q_1)-\omega_\pi(q_2)+\omega}\right\}\;.
\eea
To evaluate the principal value numerically, in an appropriate
way, one performs the following steps. First one introduces a new variable
$x\equiv \omega_\pi(q_1)$ and transforms the $q_1$-integration to an
$x$-integration, with $dq_1=\omega_\pi(q_1)/q_1dx$, and
$q_1=\sqrt{x^2-M_\pi^2}\equiv q^*$. Relabelling $q\equiv q_2$ leads to
\bea
[\Re\,\Sigma(\omega,k)]_2
&=&\left(\frac{4\lambda \sigma}{N}\right)^2
\frac{N}{8\pi^2}\frac{1}{k}
\mbox{P}\int_0^{\infty}     \, d q\,q\,[\omega_\pi(q)]^{-1}dx
\,\Theta\;\left(|q^*-q|\leq k\leq q^*+q\right)\\
&\times& \left\{ 
\frac{f(x)+f[\omega_\pi(q)]}
{x+\omega_\pi(q)-\omega}
+\frac{f(x)+f[\omega_\pi(q)]}
{x+\omega_\pi(q)+\omega}\right.
\left.+\frac{-f(x)+f[\omega_\pi(q)]}
{x-\omega_\pi(q)-\omega}
+\frac{-f(x)+f[\omega_\pi(q)]}
{x-\omega_\pi(q)+\omega}\right\}.\nn
\eea
Second one transforms the $x$-integration to a sum over $x_i$, and uses
the mean value theorem to put the terms with the distribution functions
in front of the integrals
\bea
[\Re\,\Sigma(\omega,k)]_2
&=&\left(\frac{4\lambda \sigma}{N}\right)^2
\frac{N}{8\pi^2}\frac{1}{k}
\mbox{P}\int_0^{\infty}     \, d q\,q
\, [\omega_\pi(q)]^{-1} \sum_{i}
\Theta\;\left(|q^*-q|\leq k\leq q^*+q\right)\nn\\
&\times&\left(
\{f(\hat{x})+f[\omega_\pi(q)]\}
\int_{x_i}^{x_{i+1}}dx 
\frac{1}
{x+\omega_\pi(q)-\omega}\right.\nn\\
&&+\{f(\hat{x})+f[\omega_\pi(q)]\}\int_{x_i}^{x_{i+1}}dx 
\frac{1}
{x+\omega_\pi(q)+\omega}\nn\\
&&+\{-f(\hat{x})+f[\omega_\pi(q)]\}\int_{x_i}^{x_{i+1}}dx 
\frac{1}
{x-\omega_\pi(q)-\omega}
\nn\\
&&+\left.\{-f(\hat{x})+f[\omega_\pi(q)]\}\int_{x_i}^{x_{i+1}}dx 
\frac{1}
{x-\omega_\pi(q)+\omega}\right),
\eea
where $x_i\le \hat{x}\le x_{i+1}$, and $q^*\equiv\sqrt{\hat{x}-M_\pi^2}$.
Finally, the $x$-integrals can be performed analytically, 
\bea\label{re2k}
[\Re\,\Sigma(\omega,k)]_2
&=&\left(\frac{4\lambda \sigma}{N}\right)^2
\frac{N}{8\pi^2}\frac{1}{k}
\int_0^{\infty}     \, d q\,q
\, [\omega_\pi(q)]^{-1} \sum_{i}
\Theta\;\left(|q^*-q|\leq k\leq q^*+q\right)\nn\\
&\times& \left(
\{f(\hat{x})+f[\omega_\pi(q)]\}
\ln\left|\frac{[\omega_\pi(q)+x_{i+1}]^2-\omega^2}
{[\omega_\pi(q)+x_{i}]^2-\omega^2}\right|
\right.\nn\\
&&+\left.\{-f(\hat{x})+f[\omega_\pi(q)]\}
\ln\left|\frac{[\omega_\pi(q)-x_{i+1}]^2-\omega^2}
{[\omega_\pi(q)-x_{i}]^2-\omega^2}\right|\right).
\eea
In the limit $k\to 0$, 
using Eq.~(\ref{theta_delta}), the $q$-integration 
can be carried out, 
\bea\label{re2k0}
[\Re\,\Sigma(\omega,0)]_2
&=&\left(\frac{4\lambda \sigma}{N}\right)^2
\frac{N}{4\pi^2}\sum_{i}
\,q^*\, [\omega_\pi(q^*)]^{-1}\nn\\
&\times& 
\left(
\{f(\hat{x})+f[\omega_\pi(q^*)]\}
\ln\left|\frac{[\omega_\pi(q^*)+x_{i+1}]^2-\omega^2}
{[\omega_\pi(q^*)+x_{i}]^2-\omega^2}\right|
\right.\nn\\
&&+\left.\{-f(\hat{x})+f[\omega_\pi(q^*)]\}
\ln\left|\frac{[\omega_\pi(q^*)-x_{i+1}]^2-\omega^2}
{[\omega_\pi(q^*)-x_{i}]^2-\omega^2}\right|\right).
\eea

\bibliography{largen}

\end{document}